\documentclass[prl,10pt,twocolumn]{revtex4}

\usepackage{amsmath}
\usepackage{amsfonts}
\usepackage{amssymb}
\usepackage{dcolumn}
\usepackage[cp1251]{inputenc}
\usepackage[english]{babel}
\usepackage{epsfig}
\usepackage{graphics}

\begin{document}

\title{Three-body bound states in atomic mixtures with resonant p-wave interaction}

\author{Maxim A.\ Efremov,$^{1,2}$ Lev Plimak,$^{1,3}$ Misha Yu. Ivanov,$^{3}$ and Wolfgang P.\ Schleich$^{1}$}
\affiliation{$^1$Institut f\"ur Quantenphysik and Center for
Integrated Quantum Science and Technology ($\it IQ^{ST}$),
Universit\"at Ulm, 89081 Ulm, Germany \\
$^2$A.M. Prokhorov General Physics Institute,
Russian Academy of Sciences, 119991 Moscow, Russia \\
$^3$Max-Born-Institut, 12489 Berlin, Germany}



\begin{abstract}
We employ the Born-Oppenheimer approximation to find the effective potential
in a three-body system consisting of a light particle and two heavy ones
when the heavy-light short-range interaction potential
has a resonance corresponding to a non-zero orbital angular momentum.
In the case of an exact resonance in the $p$-wave scattering amplitude, the effective potential
is attractive and long-range, namely it decreases as the third power of the inter-atomic distance.
Moreover, we show that the range and power of the potential, as well as the number of bound states are determined 
by the mass ratio of the particles and the parameters of the heavy-light short-range potential.
\end{abstract}


\maketitle

\noindent{\it Introduction.}
One of the most intriguing phenomenon of few-body physics is the Efimov effect \cite{Efimov}, which
manifests itself in an infinite number of weakly bound three-body states if at least two of the three
two-body subsystems exhibit a single weakly $s$-wave bound state or resonance.
The underlying effective potential is attractive and decreases as the {\it second} power of the inter-atomic distance \cite{Efimov}.
In this Letter, we consider a three-body system consisting
of a light particle and two heavy ones, when the heavy-light short-range interaction potential
has a weakly bound or quasi-bound, {\it i.e.}, resonant state with a non-zero orbital angular momentum.
We show that in the case of the exact $p$-wave resonance, the effective potential is also attractive and of long-range,
but now decreases as the {\it third} power of the inter-atomic distance \cite{our comment}.

The Efimov effect occurs in systems where the two-body scattering length $a_0$ is large compared to
the characteristic radius $R_0$ of the two-body interaction
and displays an universal behavior, that is the details of the underlying
short-range physics become irrelevant. 
Examples of systems with a large scattering length are halo nuclei \cite{Jensen, Nielsen-Fedorov}
and the helium trimer \cite{He-experiments}. In both cases $a_0$ is exceedingly large, but not tunable.
However, in order to observe the Efimov spectrum, it is crucial to be able to tune $a_0$.
In the domain of ultracold atomic gases this task is achieved by Feshbach resonances \cite{Feshbach-review} and
different features of the three-body recombination process as well as the scattering of the atom off the shallow dimer,
which are associated with the Efimov effect, have been measured \cite{Ferlaino-review} in this way.
An additional prerequisite to detect many Efimov states is the use of an atomic mixture \cite{rb-li}
with heavy atoms of mass $M$ and light ones of mass $m$, since in this case the ratio of two neighboring bound-state energies is
$|E_{n+1}/E_{n}|\simeq 1$ for $M/m\rightarrow \infty$ \cite{Braaten-Hammer, Fonseca, Efremov}.

In our Letter we consider such an atomic mixture. However, in contrast to the standard Efimov scenario
we focus on an exact $p$-wave resonance in the heavy-light short-range potential and determine the effective interaction potential
between the two heavy atoms. Moreover, we demonstrate that the spectrum of bound states is solely determined by the mass ratio of the heavy
and light particles, and the width of the $p$-wave state. 

Experimentally, such a mixture with a $p$-wave resonance has already been realized 
with ${\rm K}$ and ${\rm Rb}$ \cite{P-wave-mixture-KRb}, as well as with ${\rm Li}$ and ${\rm Rb}$ \cite{P-wave-mixture-LiRb}, 
corresponding to the mass ratios $m_{\rm K}/M_{\rm Rb}\approx 0.5$ and $m_{\rm Li}/M_{\rm Rb}\approx 0.1$. Therefore, 
both mixtures are promising candidates to verify our predictions.

The Efimov effect is intimately connected to the absence \cite{Braaten-Hammer} of any characteristic length scale. 
Since for a $p$-wave resonant interaction there is a natural non-zero effective range, 
there cannot \cite{Hammer-pwave,Nishida} be an Efimov effect in this case. Nevertheless, we find a series of three-body bound states. 
This surprising fact is a consequence of this non-zero effective range.

\noindent{\it Born-Oppenheimer approach.}
Our three-body system consists of a light particle which interacts with two heavy particles
and therefore can be easily analyzed within the Born-Oppenheimer approximation \cite{LL}. 
For this reason, the Schr\"odinger equation
for the full wave function $\Phi({\bf r},{\bf R}) = \Psi({\bf r};{\bf R})\chi({\bf R})$
separates into two equations and the one for the light particle reads
\begin{equation}
 \label{Schr}
    \left[-\frac{\hbar^2}{2\mu}\Delta_{\bf r}+
    U({\bf r}_{-})+U({\bf r}_{+})\right]\Psi({\bf r};{\bf R})=-\frac{\hbar^2\kappa^2}{2\mu}\Psi({\bf r};{\bf R})
\end{equation}
with ${\bf r}_{\pm}\equiv {\bf r}\pm\frac{1}{2}{\bf R}$. Here $\mu\equiv 2mM/(2M + m)\approx m$
and ${\bf R}$ denote the reduced mass and the separation between the two heavy particles, respectively.
For the sake of simplicity we assume the potential $U$ to be spherically symmetric, {\it i.e.}, $U({\bf r})=U(r)$,
and to have the finite range $R_0$, {\it i.e.}, $U(r>R_0)=0$.

The bound-state energies
\begin{equation}
 \label{V-eff}
    {\mathcal V}({\bf R})\equiv -\frac{[\hbar\kappa({\bf R)}]^2}{2\mu}
\end{equation}
of the light particle, corresponding to different expressions for $\kappa$ following from Eq. (\ref{Schr}), 
serve as effective interaction potentials for the relative motion of the heavy particles given by
\begin{equation}
 \label{Schrodinger heavy}
    \left\{\Delta_{\bf R}+\frac{M}{\hbar^2}\left[E-{\mathcal V}({\bf R})-V_0({\bf R})\right]\right\}\chi({\bf R})=0,
\end{equation}
where $E$ and $V_0$ are the total three-body energy and the direct heavy-heavy interaction potential \cite{Fonseca}, respectively. 

For atoms  $V_0$ is typically a short-range potential, and has for large distances a van-der-Waals tail, $V_0\sim 1/R^6$.  
We now show that in the case of an exact $p$-wave resonance, the effective potential ${\mathcal V}$ is a long-range one, 
${\mathcal V}\sim 1/R^3$, and therefore $V_0$ has no effect on the behavior of the total potential for large distances.

\noindent{\it Interaction potential from scattering approach.}
Next we determine ${\mathcal V}$ by a self-consistent scattering of the light particle off the two heavy ones \cite{our comment}.
For this purpose, we cast Eq. (\ref{Schr}) into the integral equation \cite{LL}
\begin{equation}
 \label{Schr-integral}
    \Psi({\bf r})=-\frac{\mu}{2\pi\hbar^2}\int d{\bf r}' \left[U({\bf r}_{-}')+U({\bf r}_{+}')\right]\Psi({\bf r}')
    \frac{e^{-\kappa |{\bf r}-{\bf r}'|}}{|{\bf r}-{\bf r}'|}\;.
\end{equation}
Since the total heavy-light potential $U({\bf r}_{-})+U({\bf r}_{+})$
is nonzero only inside two spheres of radius $R_0$ centered at
${\bf r}=\pm\frac{1}{2}{\bf R}$, we represent Eq. (\ref{Schr-integral}) as the superposition
\begin{equation}
  \label{wave-function-sum}
    \Psi({\bf r})=\Psi^{(-)}({\bf r})+ \Psi^{(+)}({\bf r})
\end{equation}
of the two waves
\begin{equation}
 \label{wave-functions}
    \Psi^{(\pm)}({\bf r})\equiv\int\limits_{|{\bf r}'\pm\frac{{\bf R}}{2}|\leq R_0}d{\bf r}'
    \sigma ^{(\pm)}({\bf r}')\frac{e^{-\kappa |{\bf r}-{\bf r}'|}}{|{\bf r}-{\bf r}'|}
\end{equation}
with
\begin{equation}
 \label{source-definition}
   \sigma ^{(\pm)}({\bf r})\equiv -\frac{\mu}{2\pi\hbar^2}\,U\left({\bf r}\pm\frac{1}{2}{\bf R}\right)\Psi({\bf r}).
\end{equation}

The expansion
\begin{equation}
 \label{exp-Green}
   \frac{e^{-\kappa |{\bf r}-{\bf r}'|}}{\kappa |{\bf r}-{\bf r}'|}=
   8\sum_{l=0}^{\infty}\sum_{|m_l|\leq l}{\mathcal{I}}_{l}(\kappa r'){\mathcal{K}}_{l}(\kappa r)
   Y_{lm_l}({\bf n_{r'}})Y_{lm_l}({\bf n_r})
\end{equation}
into the modified spherical Bessel functions ${\mathcal{I}}_{l}(z)\equiv\sqrt{\pi/(2z)}\,I_{l+1/2}(z)$ and
${\mathcal{K}}_{l}(z)\equiv\sqrt{\pi/(2z)}\,K_{l+1/2}(z)$ \cite{Abramowitz},
which is valid for $r>r'$, transforms Eq. (\ref{wave-functions}) into
\begin{equation}
 \label{wave-function-expension}
   \Psi^{(\pm)}({\bf r})=\sum_{l=0}^{\infty}\sum_{|m_l|\leq l}C_{lm_l}^{(\pm)}
    {\mathcal{K}}_{l}(\kappa r_{\pm})Y_{lm_l}({\bf n_{r_\pm}}).
\end{equation}
Here $Y_{lm_l}({\bf n_r})\equiv Y_{lm_l}(\theta_r,\varphi_r)$
are the spherical harmonics with ${\bf n_r}\equiv{\bf r}/r=\left(\theta_{\bf r},\varphi_{\bf r}\right)$.

We regard the coefficients $C_{lm_l}^{(\pm)}$ determined by the integral in Eq. (\ref{wave-functions}) as independent variables
and apply scattering theory to obtain from Eq. (\ref{wave-function-expension}) 
explicit equations for $C_{lm_l}^{(\pm)}$ coupled by the $S$-matrix elements of the potential $U$ \cite{our comment}.
For this purpose we consider a vicinity of the first potential well, that is ${\bf r}=-\frac{1}{2}{\bf R}+{\bf x}$
with $|{\bf x}|\approx R_0$, where the total solution
\begin{equation}
 \label{wave-function-expension-R0}
    \Psi\left(-\frac{{\bf R}}{2}+{\bf x}\right)=\sum_{l=0}^{\infty}\sum_{|m_l|\leq l}R_{lm_l}(\kappa|{\bf x}|)
    Y_{lm_l}({\bf n_x})
\end{equation}
given by Eq. (\ref{wave-function-sum}) can be expanded into the spherical harmonics.
Here the radial wave function
\begin{eqnarray}
 \label{solution-expansion-first sphere2}
    R_{lm_l}(\kappa|{\bf x}|)&=&C_{lm_l}^{(+)}{\mathcal{K}}_{l}(\kappa |{\bf x}|)+\nonumber \\
    & &\pi(-1)^l{\mathcal{I}}_{l}(\kappa |{\bf x}|)\sum_{l'=0}^{\infty}{\mathfrak A}_{ll'}^{(m_l)}C_{l'm_l}^{(-)}
\end{eqnarray}
is determined by the sum of the two contributions resulting from $\Psi^{(\pm)}(-\frac{1}{2}{\bf R}+{\bf x})$
defined by Eq. (\ref{wave-function-expension}), and the coefficients
\begin{eqnarray}
 \label{matrix-d-result}
    {\mathfrak A}_{ll'}^{(m_l)}(\kappa R)&\equiv&\frac{1}{\pi}\sqrt{\frac{2l+1}{2l'+1}}\sum_{L=0}^{\infty}(-1)^{L}(2L+1)\times\nonumber \\
    & &C_{l0 L0}^{l'0}C_{lm_l L0}^{l'm_l}{\mathcal{K}}_{L}(\kappa R)
\end{eqnarray}
originate from the re-expansion \cite{Varshalovich} of ${\mathcal K}_l(\kappa|{\bf x}-{\bf R}|)Y_{lm_l}({\bf n_{x-R}})$
into $Y_{lm_l}({\bf n_x})$ with the Clebsch-Gordan coefficients $C_{lm_l L0}^{l'm_l}$.

In order to derive an equation for $C_{lm_l}^{(\pm)}$ we cast the
radial wave $R_{lm_l}$ given by Eq. (\ref{solution-expansion-first sphere2})
into the superposition
\begin{equation}
 \label{solution-expansion-waves}
    R_{lm_l}(\kappa|{\bf x}|)=a_{l}(\kappa)h^{(1)}_{l}(i\kappa |{\bf x}|)+
    b_{l}(\kappa)h^{(2)}_{l}(i\kappa |{\bf x}|)
\end{equation}
of {\it outgoing} and {\it incoming} radial waves $h^{(1)}_{l}$ and $h^{(2)}_{l}$ with amplitudes
\begin{equation}
 \label{solution-expansion-amplitudes-a}
  a_{l}(\kappa)=-\frac{\pi i^l}{2} C_{lm_l}^{(+)}+\frac{\pi i^l}{2}
  \sum_{l'=0}^{\infty}{\mathfrak A}_{ll'}^{(m_l)}(\kappa R)C_{l'm_l}^{(-)}
\end{equation}
and
\begin{equation}
 \label{solution-expansion-amplitudes-b}
  b_{l}(\kappa)=\frac{\pi i^l}{2}\sum_{l'=0}^{\infty}{\mathfrak A}_{ll'}^{(m_l)}(\kappa R)C_{l'm_l}^{(-)}.
\end{equation}
The spherical Bessel functions of the third kind
$h^{(1)}_{l}$ and $h^{(2)}_{l}$ are determined \cite{Abramowitz} in terms of ${\mathcal{K}}_{l}$ and ${\mathcal{I}}_{l}$
as ${\mathcal{K}}_l(z)=-(\pi i^l/2) h_{l}^{(1)}(iz)$ and ${\mathcal{I}}_l(z)=[h_l^{(1)}(iz)+h_l^{(2)}(iz)]/(2i^l)$.

Since the amplitudes $a_l$ and $b_l$ of the {\it outgoing} and {\it incoming} waves
are coupled \cite{LL} by the $S$-matrix elements $S_l$ of the scattering potential $U$, that is
\begin{equation}
 \label{scattering matrix}
    a_l(\kappa)=S_l(i\kappa)b_l(\kappa),
\end{equation}
we arrive at
\begin{equation}
 \label{system-final1}
    C_{lm_l}^{(+)}+\left[S_l(i\kappa)-1\right]
    \sum_{l'=0}^{\infty}{\mathfrak A}_{ll'}^{(m_l)}(\kappa R)C_{l'm_l}^{(-)}=0.
\end{equation}

Similarly we obtain from the second potential well, centered at ${\bf r}=\frac{1}{2}{\bf R}$, the relation
\begin{equation}
 \label{system-final2}
    C_{lm_l}^{(-)}+\left[S_l(i\kappa)-1\right]
    \sum_{l'=0}^{\infty}(-1)^{l+l'}{\mathfrak A}_{ll'}^{(m_l)}(\kappa R)C_{l'm_l}^{(+)}=0.
\end{equation}

Equations (\ref{system-final1}) and (\ref{system-final2}) constitute
a system of linear algebraic equations for $C_{lm_l}^{(\pm)}$
determining via Eq. (\ref{wave-function-expension}) the waves $\Psi^{(\pm)}$.
Its solution is nonzero only if the corresponding determinant vanishes which provides us with
a transcendental equation for $\kappa=\kappa(R)$, and thus for
the interaction potential ${\mathcal V}$ defined by Eq. (\ref{V-eff}).
The coefficients of these equations are determined by the $S$-matrix elements
of the interaction potential $U$ between the heavy and the light atoms.

\noindent{\it Zero-range limit.}
In order to test our method, we first consider a zero-range potential,
for which only $s$-wave scattering occurs and the $S$-matrix elements read \cite{Demkov}
\begin{equation}
 \label{zero-range}
    S_l(i\kappa)-1=\frac{2\kappa}{1/a_0-\kappa}\;\delta_{l,0}.
\end{equation}
In this case, the system Eqs. (\ref{system-final1}) and (\ref{system-final2})
reduces to two algebraic equations for $C_{00}^{(\pm)}$ and has non-trivial solutions only if
\begin{equation}
 \label{energy-eq-s0}
   \left[S_0(i\kappa)-1\right]{\mathfrak A}_{00}^{(0)}(\kappa R)=\pm 1,
\end{equation}
with ${\mathfrak A}_{00}^{(0)}(\kappa R)\equiv[1/(2\kappa R)]e^{-\kappa R}$
defined by Eq. (\ref{matrix-d-result}). This condition translates into equation
\begin{equation}
 \label{energy-eq-Efimov}
    \frac{1}{\xi-\alpha_0\rho}\,e^{-\xi}=\pm 1
\end{equation}
for $\xi\equiv\kappa R$ with the parameters
\begin{equation}
 \label{a1-rho-parameters}
    \alpha_0\equiv \frac{R_0}{a_0}\;\;\;{\rm and}\;\;\;\rho\equiv\frac{R}{R_0},
\end{equation}
and coincides with the equation for the bound-state energy obtained in Refs. \cite{Fonseca, Demkov} for the
case of the zero-range potential.

In the case of a $s$-wave resonance, that is $\alpha_0=0$, Eq. (\ref{energy-eq-Efimov})
has a solution $\xi=\xi_* \approx 0.57$ only for the plus sign on the right-hand side,
which translates into the familiar Efimov potential
\begin{equation}
 \label{energy-Efimov-asymptotic}
    {\mathcal V}^{(0)}\equiv -\frac{[\hbar\kappa_{+}(R)]^2}{2\mu}= -\frac{\hbar^2}{2\mu}\frac{\xi_*^2}{R^2},
\end{equation}
decaying with the second power of $R$.

\noindent{\it P-wave resonance.}
Next we focus on the low-energy limit, that is on $|E|\ll \hbar^2/(\mu R_0^2)$,
or $\kappa R_0\ll 1$, where the $S$-matrix elements 
\begin{equation}
 \label{S-matrix-phase}
  S_l(i\kappa)-1=\frac{2i}{\cot[\delta_l(i\kappa)]-i}
\end{equation}
in the $l$-th partial wave are determined \cite{LL} by the scattering phases $\delta_l(i\kappa)$ following \cite{Mott}
from the effective-range expansion
\begin{equation}
 \label{S-matrix-low-energy}
  (i\kappa)^{2l+1}\cot[\delta_l(i\kappa)]\cong -1/a_l+(r_{l}/2)\kappa^2.
\end{equation}

A resonance in the $l$-th partial wave is reached when the absolute value of
the {\it effective scattering length} $|a_l|\gg R_0^{2l+1}$. 
The effective range $r_l$ for $l>0$ is positive  and linked
\cite{LL, Feshbach-p-wave, Feshbach-comment} to the width of the resonance. Moreover, for any short-range potential 
$r_l$ has a lower bound \cite{Feshbach-p-wave}, that is $r_l\geq\tilde{\alpha}_l R_0^{1-2l}$ 
with the positive constant $\tilde{\alpha_l}$ determined by $l$.

Recently it has been shown \cite{Hammer-pwave} that in the unitary limit, defined as $1/a_1=0$ and $r_1=0$, 
any characteristic length scale disappears and the three-body problem could exhibit the Efimov effect. 
However, due to the natural lower bound \cite{Feshbach-p-wave} on $r_1$, this limit is unphysical and eliminates \cite{Nishida} 
the possibility of the Efimov effect in our system.         
We now take into account this lower bound on $r_1$ and 
show that the effective potential, and the corresponding energy spectrum are induced by and depend explicitly on $r_1$ \cite{our comment}.   

For this purpose we consider the case of a resonant $p$-wave, that is a partial wave with $l=1$, and substitute
the $S$-matrix elements given by Eqs. (\ref{S-matrix-phase}) and (\ref{S-matrix-low-energy}) into Eqs.
(\ref{system-final1}) and (\ref{system-final2}).
Since $S_0(i\kappa)-1\sim S_1(i\kappa)-1 \sim \kappa R_0$ and $S_{l>1}(i\kappa)-1\sim (\kappa R_0)^{2l+1}$,
we neglect the small terms with $l=2,3,...$ and arrive at two separate systems of equations
for each projection $m_l$ of the angular momentum.

For $m_l=\pm 1$, Eqs. (\ref{system-final1}) and (\ref{system-final2})
simplify to two equations for $C_{1,\pm 1}^{(\pm)}$ and
have a non-trivial solution only if
\begin{equation}
 \label{energy-eq-m1-first}
    \left[S_1(i\kappa)-1\right]{\mathfrak A}_{11}^{(\pm 1)}(\kappa R)=\pm 1.
\end{equation}
Since $S_1(i\kappa)$ is given by Eqs. (\ref{S-matrix-phase}) and (\ref{S-matrix-low-energy}) and
${\mathfrak A}_{11}^{(\pm 1)}(\kappa R)\equiv -\frac{3}{2}[(1+\kappa R)/(\kappa R)^3]e^{-\kappa R}$,
Eq. (\ref{energy-eq-m1-first}) for $\xi=\kappa^{(1)}R$ and distances $R\geq 2R_0$ reads
\begin{equation}
 \label{energy-eq-m1}
    \frac{(1+\xi)}{\beta\rho\xi^2-\alpha_1\rho^3-\frac{1}{3}\xi^3}\,e^{-\xi}=\pm 1
\end{equation}
with the dimensionless parameters
\begin{equation}
 \label{alpha-beta-parameters}
    \alpha_1\equiv\frac{R_0^3}{3a_1}\;\;\;{\rm and}\;\;\;\beta\equiv\frac{r_1R_0}{6}\,.
\end{equation}

In the resonant case, that is $\alpha_1=0$, Eq. (\ref{energy-eq-m1})
has a solution only for the plus sign on the right-hand side.
In the limit of $0<\xi\ll 1$, we find $\xi\cong (\beta\rho)^{-\frac{1}{2}}$ for $\rho\gg 1$,
which translates into the potential
\begin{equation}
 \label{energy-m1-asymptotic}
    {\mathcal V}^{(1,\pm 1)}\equiv -\frac{[\hbar\kappa_{+}^{(1)}(R;0)]^2}{2\mu}\cong
    -\frac{\hbar^2}{2\mu}\frac{6}{r_1R^3},
\end{equation}
which is independent of $R_0$.



The form of the potentials ${\mathcal V}_{\pm}^{(1,\pm 1)}(R;\alpha_1)\equiv -[\hbar^2/(2\mu R_0^2)](\xi_{\pm}/\rho)^2$ 
is determined by the solutions $\xi_{\pm}(\rho;\alpha_1)$ of Eq. (\ref{energy-eq-m1}) and depends on the sign of $a_1$,
that is on the sign of $\alpha_1$, Eq. (\ref{alpha-beta-parameters}).
Indeed, for $\alpha_1>0$, {\it i.e.}, in the case of the weakly-bound $p$-wave state in $U$,
${\mathcal V}_{+}^{(1,\pm 1)}$ as well as ${\mathcal V}_{-}^{(1,\pm 1)}$ approach for large distances, $R>R_0 |\alpha_1|^{-1/3}$,
the bound state energy $\varepsilon_1\equiv -(\alpha_1/\beta)$ of the light particle.
For short distances, $R<R_0 |\alpha_1|^{-1/3}$, ${\mathcal V}_{+}^{(1,\pm 1)}(R;\pm|\alpha_1|)$ approach
${\mathcal V}_{+}^{(1,\pm 1)}(R;\alpha_1=0)={\mathcal V}^{(1,\pm 1)}(R)$.

In the case of a $p$-wave resonance, the matrix element $S_1$
corresponding to the resonant channel is of the same order as $S_{0}$
for the non-resonant channel \cite{LL}. Therefore, for $m_l=0$
we have to take into account in Eqs. (\ref{system-final1}) and (\ref{system-final2}) both the $s$- and $p$-waves,
which gives rise to a system of four algebraic equations for $C_{0,0}^{(\pm)}$ and $C_{1,0}^{(\pm)}$, 
leading us to the relation
\begin{eqnarray}
 \label{energy-eq-m0-first}
    1-(S_0-1)(S_1-1)
    \left({\mathfrak A}_{00}^{(0)}{\mathfrak A}_{11}^{(0)}-{\mathfrak A}_{01}^{(0)}{\mathfrak A}_{10}^{(0)}\right)=\nonumber \\
    \mp\left[(S_0-1){\mathfrak A}_{00}^{(0)}-(S_1-1){\mathfrak A}_{11}^{(0)}\right].
\end{eqnarray}
According to Eqs. (\ref{matrix-d-result}), (\ref{S-matrix-phase}), (\ref{S-matrix-low-energy}) and (\ref{alpha-beta-parameters}), we obtain
${\mathfrak A}_{01}^{(0)}\equiv{\mathfrak A}_{10}^{(0)}=-\frac{\sqrt{3}}{2}[(1+\kappa R)/(\kappa R)^2]e^{-\kappa R}$ and
${\mathfrak A}_{11}^{(0)}\equiv\frac{3}{2}[(\kappa^2 R^2+2\kappa R+2)/(\kappa R)^3]e^{-\kappa R}$,
and Eq. (\ref{energy-eq-m0-first}) for $\xi=\kappa^{(0)}R$ takes the form
\begin{eqnarray}
 \label{energy-eq-m0}
    1+\frac{e^{-2\xi}}{(\xi-\alpha_0\rho)
    (\beta\rho\xi^2-\alpha_1\rho^3-\frac{1}{3}\xi^3)}=\nonumber \\
    \pm e^{-\xi}\left[\frac{(\xi^2+2\xi+2)}{\beta\rho\xi^2-\alpha_1\rho^3-\frac{1}{3}\xi^3}
    +\frac{1}{\xi-\alpha_0\rho}\right].
\end{eqnarray}
In the resonant case, $\alpha_1=0$, we have $\alpha_0\sim 1$ and Eq. (\ref{energy-eq-m0})
has a solution only for the plus sign on the right-hand side.
In the limit of $0<\xi\ll 1$, we find $\xi\cong [2/(\beta\rho)]^{\frac{1}{2}}$ for $\rho\gg 1$,
giving rise to the potential
\begin{equation}
 \label{energy-m0-asymptotic}
    {\mathcal V}^{(1,0)}\equiv -\frac{[\hbar\kappa_{+}^{(0)}(R; 0)]^2}{2\mu}\cong
    -\frac{\hbar^2}{\mu}\frac{6}{r_1 R^3}.
\end{equation}

The potentials ${\mathcal V}_{\pm}^{(1,0)}(R;\alpha_1)\equiv -[\hbar^2/(2\mu R_0^2)](\xi_{\pm}/\rho)^2$
determined by the solutions $\xi_{\pm}(\rho;\alpha_1)$ of Eq. (\ref{energy-eq-m0}) at $\alpha_0\sim 1$ and $\beta\sim 1$
are similar to ${\mathcal V}_{\pm}^{(1,\pm 1)}$ with two qualitative differences:
(i) the ranges of ${\mathcal V}_{\pm}^{(1,\pm 1)}$ and ${\mathcal V}_{\pm}^{(1,0)}$
are different and equal to $R_1\equiv R_0|\alpha_1|^{-1/3}$ and $R_2\equiv R_0(0.5\,|\alpha_1|)^{-1/3}$, respectively, and
(ii) in the case of exact resonance, ${\mathcal V}^{(1,0)}$ given by Eq. (\ref{energy-m0-asymptotic}) has the same asymptotic behavior
as ${\mathcal V}^{(1,\pm 1)}$ defined by Eq. (\ref{energy-m1-asymptotic}) with twice the amplitude.

\noindent{\it Spectrum of induced $1/R^3$-potential.} Finally we
focus on the dynamics of the two heavy particles dictated by the
Schr\"odinger equation (\ref{Schrodinger heavy}) with the potential
${\mathcal V}$ given by Eqs. (\ref{energy-m1-asymptotic}) and
(\ref{energy-m0-asymptotic}) and induced by the $p$-wave resonance
in the light-heavy interaction \cite{our comment}. We emphasize that ${\mathcal V}$ is
only meaningful for $R\gg R_0$, since for $R\sim R_0$ 
the dynamics is determined by the direct short-range forces.

The energies $E_n$ of the bound states with zero angular
orbital momentum follow from the familiar WKB quantization rule \cite{LL}
\begin{equation}
 \label{qauntization dimensionless}
    n-n_0=\frac{1}{\pi\hbar}\int\limits_{R_0}^{R_{E_n}}\sqrt{M[E_n-{\mathcal V}(R)]}\,dR,
\end{equation}
giving rise \cite{WKB, Mueller-Friedrich} to the spectrum
\begin{equation}
 \label{qauntization}
    E_n=-\frac{\hbar^2}{MR_*^2}\left(\frac{n_0-n}{g}\right)^6
\end{equation}
for the weakly bound states, induced exclusively by $1/R^3$-potential. 
Here we have introduced the characteristic range 
\begin{equation}
 \label{effective range}
    R_* \equiv\left(2-|m_l|\right)\frac{3M}{\mu r_1}
\end{equation}
of the effective potential ${\mathcal V}$ in the resonant case, that is $\alpha_1=0$, and 
$[n_0]-n=1,2,..$, where the integer part $[n_0]$, determined by the phase of the wave function at the short
distances $R\sim R_0$, plays a role of a three-body parameter and 
$g\equiv\Gamma(\frac{5}{6})/[\sqrt{\pi}\,\Gamma(\frac{4}{3})]$.

Since ${\mathcal V}$ defined by Eqs. (\ref{energy-m1-asymptotic})
and (\ref{energy-m0-asymptotic}) has a tail falling off faster than
$-1/R^2$, it supports \cite{LL} only a finite number $N_0$ of bound
states with zero angular orbital moment. 
Indeed, $N_0$ can be estimated \cite{LL} by the WKB method and yields 
\begin{equation}
 \label{N bound states}
    N_0=\frac{1}{\pi}\int\limits_{R_0}^{\infty}\sqrt{\frac{R_*}{R^3}}\,dR
    =\frac{2}{\pi}\sqrt{\frac{R_*}{R_0}}=
    \frac{2}{\pi}\left[\frac{(2-|m_l|)}{r_1R_0}\frac{3M}{m}\right]^{\frac{1}{2}},
\end{equation}
that is $N_0$ is determined by the square root of the ratio of the mass-ratio
$M/m$ to the dimensionless effective range $r_1R_0$ of the $p$-wave
resonance.

The appearance of ${\mathcal V}$ given by Eqs. (\ref{energy-m1-asymptotic}) and (\ref{energy-m0-asymptotic})
can be verified experimentally by scattering a heavy atom off the diatomic molecule consisting of a heavy and a light atom.
The predicted three-body bound states manifest themselves as resonances in the cross-section
of the atom-molecule scattering when we tune the magnetic field close to the $p$-wave Feshbach resonance.
Moreover, due to the inverse-cube tail the cross-section $\sigma_L$ of the $L$-th partial wave has the unique behavior
\cite{Gao}, $\sigma_0(E)=\pi R_*^2 \ln^2(MR_*^2E/\hbar^2)$ and $\sigma_{L>0}(E)=\pi R_*^2[(2L+1)/(L^2+L)^2]$ at the low incident energy
$E\ll \hbar^2/(MR_*^2)$.

\noindent{\it Summary and outlook.} We have found a novel series of bound states in the three-body system consisting of a light particle
and two heavy ones when the heavy-light short-range interaction potential has the $p$-wave resonance.
In the case of an exact resonance, the effective potential is attractive and of long-range.
Moreover, the spectrum of bound states is determined by the mass ratio of the heavy and light particles
as well as the parameters of the heavy-light short-range potential.

Our treatment is based on the Born-Oppenheimer approximation, which for the Efimov case has already been shown  
to provide the correct effective interaction potential \cite{Hahn} and is not limited \cite{Helfrich} to large mass ratios.
Therefore, we are confident that our approach is also adequate for the three-body system with a $p$-wave resonant interaction.

\noindent{\it Acknowledgments.}
We are deeply indebted to F. Ferlaino, K. Fossez, R. Kaiser, D. Petrov, M. Ploszajczak, and R. Walser for stimulating discussions.
We also appreciate the financial support by the German Science Foundation (DFG) in the framework of the SFB/TRR-21.
M.A.E. and M.Y.I. are grateful to the Alexander von Humboldt Stiftung and M.A.E. thanks the Russian Foundation for Basic Research (grant 10-02-00914-a).



\begin{thebibliography}{99}
\bibitem{Feshbach-comment} Experimentally, the resonant regime can be achieved using Feshbach resonances.
In the vicinity of a resonance in the $l$-th partial wave,
corresponding to the magnetic field $B\simeq B_l$, the effective scattering length $a_l(B)\propto 1/(B_l-B)$ and
the range $r_l(B)$ is almost constant and determined by the width of the Feshbach resonance. 
Indeed, the cases of a broad or a narrow resonance correspond to $r_1R_0\sim 1$ or $r_1R_0\gg 1$, respectively.
\end{thebibliography}
\end{document}